\documentstyle[12pt,epsf]{article}

\begin{document}

\baselineskip 6mm
\renewcommand{\thefootnote}{\fnsymbol{footnote}}


\newcommand{\nc}{\newcommand}
\newcommand{\rnc}{\renewcommand}


\rnc{\baselinestretch}{1.24}    
\setlength{\jot}{6pt}       
\rnc{\arraystretch}{1.24}   

\makeatletter
\rnc{\theequation}{\thesection.\arabic{equation}}
\@addtoreset{equation}{section}
\makeatother



\nc{\be}{\begin{equation}}

\nc{\ee}{\end{equation}}

\nc{\bea}{\begin{eqnarray}}

\nc{\eea}{\end{eqnarray}}

\nc{\xx}{\nonumber\\}

\nc{\ct}{\cite}

\nc{\la}{\label}

\nc{\eq}[1]{(\ref{#1})}

\nc{\newcaption}[1]{\centerline{\parbox{6in}{\caption{#1}}}}

\nc{\fig}[3]{

\begin{figure}
\centerline{\epsfxsize=#1\epsfbox{#2.eps}}
\newcaption{#3. \label{#2}}
\end{figure}
}


\def\CA{{\cal A}}
\def\CC{{\cal C}}
\def\CD{{\cal D}}
\def\CE{{\cal E}}
\def\CF{{\cal F}}
\def\CG{{\cal G}}
\def\CH{{\cal H}}
\def\CK{{\cal K}}
\def\CL{{\cal L}}
\def\CM{{\cal M}}
\def\CN{{\cal N}}
\def\CO{{\cal O}}
\def\CP{{\cal P}}
\def\CR{{\cal R}}
\def\CS{{\cal S}}
\def\CU{{\cal U}}
\def\CW{{\cal W}}
\def\CY{{\cal Y}}


\def\IR{{\hbox{{\rm I}\kern-.2em\hbox{\rm R}}}}
\def\IB{{\hbox{{\rm I}\kern-.2em\hbox{\rm B}}}}
\def\IN{{\hbox{{\rm I}\kern-.2em\hbox{\rm N}}}}
\def\IC{\,\,{\hbox{{\rm I}\kern-.59em\hbox{\bf C}}}}
\def\IZ{{\hbox{{\rm Z}\kern-.4em\hbox{\rm Z}}}}
\def\IP{{\hbox{{\rm I}\kern-.2em\hbox{\rm P}}}}
\def\IH{{\hbox{{\rm I}\kern-.4em\hbox{\rm H}}}}
\def\ID{{\hbox{{\rm I}\kern-.2em\hbox{\rm D}}}}


\def\a{\alpha}
\def\b{\beta}
\def\ga{\gamma}
\def\d{\delta}
\def\ep{\epsilon}
\def\ph{\phi}
\def\k{\kappa}
\def\l{\lambda}
\def\m{\mu}
\def\n{\nu}
\def\th{\theta}
\def\rh{\rho}
\def\s{\sigma}
\def\t{\tau}
\def\w{\omega}
\def\G{\Gamma}


\def\half{\frac{1}{2}}
\def\dint#1#2{\int\limits_{#1}^{#2}}
\def\goto{\rightarrow}
\def\para{\parallel}
\def\brac#1{\langle #1 \rangle}
\def\grad{\nabla}
\def\curl{\nabla\times}
\def\div{\nabla\cdot}
\def\p{\partial}
\def\e{\epsilon_0}


\def\Tr{{\rm Tr}\,}
\def\det{{\rm det}}


\def\vare{\varepsilon}
\def\barz{\bar{z}}
\def\barw{\bar{w}}


\def\ad{\dot{a}}
\def\bd{\dot{b}}
\def\cd{\dot{c}}
\def\dd{\dot{d}}
\def\so{SO(4)}
\def\sop{SO(4)^\prime}
\def\bc{{\bf C}}
\def\bfz{{\bf Z}}
\def\bz{\bar{z}}

\begin{titlepage}


\hfill\parbox{3.7cm} {SNUTP 04-007 \\
{\tt hep-th/0404064}}

\vspace{15mm}

\begin{center}
{\Large \bf Exact Seiberg-Witten Map, Induced Gravity
and Topological Invariants  in Noncommutative Field Theories}

\vspace{10mm}
Rabin Banerjee$^{\, a \,}$\footnote{rabin@bose.res.in}
and Hyun Seok Yang$^{\, b \,}$\footnote{hsyang@phya.snu.ac.kr}
\\[10mm]

${}^a${\sl S. N. Bose National Center for Basic Sciences, \\
JD Block, Sector III, Salt Lake City, Calcutta-700098, India} \\
${}^b${\sl School of Physics, Seoul National University,
Seoul 151-747, Korea} \\
\end{center}

\thispagestyle{empty}

\vskip1cm


\centerline{\bf ABSTRACT}
\vskip 4mm
\noindent

We revisit the exact Seiberg-Witten (SW) map on Dirac-Born-Infeld
actions, making a connection with the deformation quantization
scheme. The picture on field dependent induced gravity from
noncommutativity becomes more transparent in the context of
deformation quantization. We also find an exact SW map for an
adjoint scalar field, consistent with that deduced from RR
couplings of unstable non-BPS D-branes. The dual description via
the exact SW map can again be interpreted as the ordinary field
theory coupling to gravity induced by gauge fields. Using the
exact SW maps, we further discuss several aspects of topological
invariants in noncommutative (NC) gauge theory. Especially, it is
shown that the K-theory class on NC instantons is mapped to the
usual second Chern class via exact SW map and it leads to an exact
SW map between commutative and NC Chern-Simons terms.

\vspace{1cm}

\today

\end{titlepage}

\renewcommand{\thefootnote}{\arabic{footnote}}
\setcounter{footnote}{0}

\section{Introduction}

A noncommutative (NC) space is obtained by quantizing a given
space with its symplectic structure, treating it as a phase space.
Also field theories can be formulated on a NC space. NC field
theory means that fields are defined as functions over the NC
space. At the algebraic level, the fields become operators acting
on a Hilbert space as a representation space of the NC space.
Since the NC space resembles a quantized phase space, the idea of
localization in ordinary field theory is lost. The notion of a
point is replaced by that of a state in the representation space.
Thus it may help understanding non-locality at short distances in
quantum gravity.

Recently it has been known \ct{nc-space,review} that
NC field theories can arise naturally as a decoupled limit of open
string dynamics on D-branes in the background of a Neveu-Schwarz
$B$ field. The open string effective action on a D-brane is given
by the Dirac-Born-Infeld (DBI) action in the limit of slowly
varying fields \ct{dbi}. Seiberg and Witten, however, showed \ct{sw}
that an explicit form of the effective action depends on the
regularization scheme of two dimensional field theory defined by
the worldsheet action. That is, depending on the regularization
scheme or path integral prescription for the open string ending on
a D-brane, one can have two descriptions: commutative and NC
descriptions. Since these two descriptions arise from the same
open string theory depending on different regularizations and the
physics should not depend on the regularization scheme, Seiberg
and Witten argued \ct{sw} that the two descriptions should be equivalent
and thus there must be a spacetime field redefinition between
ordinary and NC gauge fields, so called Seiberg-Witten (SW) map.

In this sense NC gauge theories have a dual description through
the SW map in terms of ordinary gauge theory on commutative
spacetime. To understand the dual description exactly, it is
important to know the exact SW map between the gauge fields.
In a recent work \ct{yang}, it
was pointed out that there is an extremely simple way to find the
exact SW map using the change of variables between the open and
closed string parameters. The resulting exact SW map revealed a
remarkable picture that the NC Maxwell action can be regarded as
the ordinary Maxwell action coupling to a metric deformed by gauge
fields, which genuinely realizes an interesting idea
by Rivelles \ct{rivelles}.

Why NC gauge fields play a role of gravity may be understood by
noting \ct{ind-gravity,das-rey} that translations in the NC directions are
equivalent to a gauge transformation up to global symmetry
transformations. Based on this property the authors
in \ct{ind-gravity} assert that
NC gauge theories are toy models of general relativity.
We quote a paragraph in \ct{ind-gravity}:
\begin{quote}
What is unusual about noncommutative gauge theories is that {\it translations
in the noncommutative directions are equivalent to a combination
of a gauge transformation and a constant shift of the gauge
field.} This explains why in noncommutative gauge theories there
do not exist local gauge invariant observables, since by a gauge
transformation we can effect a spatial translation ! This is
analogous to the situation in general relativity, where
translations are also equivalent to gauge transformations (general
coordinate transformations) and one cannot construct local gauge
invariant observables. The fact that spatial translations are
equivalent to gauge transformations (up to global symmetry
transformations) is one of the most interesting features of
noncommutative gauge theories. These theories are thus toy models
of general relativity - the only other theory that shares this
property.
\end{quote}

If one employs commutative description via SW map, however, the
connection between translations and gauge transformations is lost.
A global translation on commutative fields can no longer be
rewritten as a gauge transformation. So one may wonder how the
property of NC field theories show up in the commutative
description via SW map. Now the aspect concerning gravity directly
emerges as an effective metric induced by gauge fields when the
commutative description is employed.
Indeed this was the motivation in \ct{rivelles} to
explore the connection between NC field theories and gravity.

This paper is organized as follows. In Sec. 2, we briefly
summarize the exact SW maps on DBI actions
obtained in \ct{yang} to make a connection with later sections.
In Sec. 3, we adopt the deformation quantization scheme a la
Kontsevich \ct{kon} to show that the results in \ct{yang} can be reproduced
in this approach too. The picture on the induced gravity from
noncommutativity becomes more transparent in the context of
deformation quantization. In Sec. 4, we find an exact SW map for a
scalar field in the adjoint representation of gauge group and show
that it is consistent with that deduced from RR couplings of
unstable non-BPS D-branes \ct{mukhi-sury}. The dual
description via the exact SW map can again be interpreted as the
ordinary field theory coupling to dilaton gravity induced by gauge
fields. In Sec. 5, we discuss several aspects of topological
invariants in NC gauge theory using the exact SW maps. Especially,
it is shown that the K-theory class on NC instantons is mapped to
the usual second Chern class via exact SW map and it leads to an
exact SW map between commutative and NC Chern-Simons
terms, which was proved earlier in \ct{grandi}. In Sec. 6, we
briefly summarize our results obtained and discuss some related
open issues.

\section{Exact Seiberg-Witten Map and Induced Gravity from Noncommutativity}

In this section, we recapitulate the exact SW maps on DBI
actions obtained in \ct{yang} to make a connection with later sections.
The worldsheet action governing the open string dynamics attached
on $Dp$-branes in flat spacetime, with metric $g_{\mu\nu}$, in the presence of
a constant Neveu-Schwarz $B$-field is given by
\be \la{string-action-1}
S = \frac{1}{2\kappa}\int_{\Sigma} d^2 \sigma g_{\mu\nu} \p_a
x^\mu \p^a x^\nu + i \int_{\p \Sigma} d\tau \Bigl( \half B_{\mu\nu} x^\nu
- A_\mu (x) \Bigr) \p_\tau x^\mu,
\ee
where the string worldhseet $\Sigma$ is the upper half plane parameterized by $-\infty
\leq \tau \leq \infty$ and $0 \leq \sigma \leq \infty$ and $\p
\Sigma$ is its boundary. We define the inverse string tension as
\be \la{kappa}
\kappa \equiv 2 \pi \alpha^\prime,
\ee
which is a useful expansion parameter in a low energy effective
action of D-branes. The propagator evaluated at boundary
points \ct{sw} is
\begin{equation}\label{open-propagator}
    \langle x^\mu (\tau)  x^\nu (\tau^\prime) \rangle = -
    \frac{\kappa}{2\pi} \Bigl(\frac{1}{G}\Bigr)^{\mu\nu} \log(\tau - \tau^\prime)^2 +
    \frac{i}{2}  \theta^{\mu\nu} \epsilon(\tau - \tau^\prime)
\end{equation}
where $\epsilon(\tau)$ is the step function. Here
\bea \la{open-g-inverse}
&& \Bigl(\frac{1}{G}\Bigr)^{\mu\nu} = \Bigl( \frac{1}{g + \kappa B}g \frac{1}{g - \kappa
B}\Bigr)^{\mu\nu}, \\
\la{open-g}
&& G_{\mu\nu} = g_{\mu\nu} - \kappa^2 (B g^{-1}B)_{\mu\nu}, \\
\la{open-theta}
&& \theta^{\mu\nu} = - \kappa^2 \Bigl( \frac{1}{g + \kappa B}B
\frac{1}{g - \kappa B}\Bigr)^{\mu\nu}.
\eea
In what follows, we will often use the matrix notation:
\be \label{def-matrix}
AB =  A_{\mu \alpha} B^{\alpha \mu},
    \quad (AB)_{\mu\nu} =  A_{\mu \alpha} {B^\alpha}_\nu, \quad \rm{etc}.
\ee
>From Eqs. \eq{open-g-inverse} and \eq{open-theta},
we have the following relation
\begin{equation} \label{op-cl}
\frac{1}{G} + \frac{\theta}{\kappa} =  \frac{1}{g + \kappa B}.
\ee
The object $G_{\mu\nu}$ has a simple interpretation as the
effective metric seen by the open strings while $g_{\mu\nu}$ is
the closed string metric. Furthermore the coefficient
$\theta^{\mu\nu}$ has a simple interpretation as
\begin{equation}\label{nc-space}
    [ x^\mu (\tau),  x^\nu (\tau) ] = i \theta^{\mu\nu}.
\end{equation}
That is, $x^\mu$ are coordinates on a NC space with
noncommutativity parameter $\theta$ \ct{nc-space}.

For a slowly varying approximation of neglecting
derivative terms, i.e., $ \sqrt{\kappa} |\frac{\p F}{F}| \ll 1$,
the spacetime low energy effective action on a single $Dp$-brane
is given by the DBI action \ct{dbi}
\begin{equation}\label{dbi-c}
S(g_s, g, A, B) = \frac{2\pi}{g_s (2\pi \kappa)^{\frac{p+1}{2}}}\int d^{p+1} x
\sqrt{-\det(g + \kappa (B+F))},
\end{equation}
where
\begin{equation}\label{c-f}
F_{\mu\nu} = \p_\mu A_\nu -  \p_\nu A_\mu.
\end{equation}
Note that the effective action is expressed in terms of closed
string variables $g_{\mu\nu}, B_{\mu\nu}$ and $g_s$.
Seiberg and Witten, however, showed \ct{sw} that an explicit form of the effective action
depends on the regularization scheme of two dimensional field
theory defined by the worldsheet action \eq{string-action-1},
which is related to field redefinition in spacetime.

As was explained in \ct{sw}, there is a general description with an arbitrary $\theta$
associated with a suitable regularization that interpolates
between Pauli-Villars and point-splitting. This freedom is
basically coming from the fact that the sigma model \eq{string-action-1}
has a symmetry $ A \to A + \Lambda, \; B \to B-d\Lambda $, for any
one-form $\Lambda$ and thus the open string theory depends only on the gauge invariant
combination ${\cal F} = B + F$. Given such a symmetry,
there is a freedom of shift in $B$ keeping fixed ${\cal F}$.
By taking the background to be $B$ or $B^\prime$,
we get a NC description with appropriate $\theta$ or
$\theta^\prime$, and different $F$'s. The freedom in the description is
parameterized by a two-form $\Phi$. In this case the
change of variables found by Seiberg and Witten \ct{sw} is given by
\bea \la{op-cl-gen}
&& \frac{1}{G + \kappa \Phi} + \frac{\theta}{\kappa}
=  \frac{1}{g + \kappa B}, \\
&& \label{Gs-gs-gen}
G_s = g_s \sqrt{\frac{\det (G + \kappa \Phi)}{\det (g + \kappa B)}}.
\eea
The effective action in these variables is given by
\begin{equation}\label{dbi-gen}
\widehat{S}_\Phi(G_s, G, \widehat{A}, \theta) =
\frac{2\pi}{G_s (2\pi \kappa)^{\frac{p+1}{2}}}\int d^{p+1} x
\sqrt{-\det(G + \kappa (\widehat{F}+ \Phi))}.
\end{equation}
The action depends on the open string variables $G_{\mu\nu}, \theta_{\mu\nu}$
and $G_s$, where the $\theta$-dependence is entirely in the
$\star$ product in the field strength $\widehat{F}$:
\be \la{nc-f}
\widehat{F}_{\mu\nu} = \p_\mu \widehat{A}_\nu -  \p_\nu \widehat{A}_\mu
- i \widehat{A}_\mu \star \widehat{A}_\nu + i \widehat{A}_\nu
\star \widehat{A}_\mu.
\ee

For every background characterized by $B, g_{\mu\nu}$ and $g_s$,
we thus have a continuum of descriptions labelled by a choice of
$\Phi$. Indeed, for $\Phi=B$ where $G=g, \; G_s = g_s$
and $\theta =0$, $\widehat{S}_\Phi$ recovers the commutative description
\eq{dbi-c} while $\Phi=0$ leads to the familiar NC description.
Seiberg and Witten \ct{sw} proved that DBI actions are independent
of the choice $\Phi$, namely,
\begin{equation}\label{equiv-dbi-gen}
 \widehat{S}_\Phi (G_s, G, \widehat{A}, \theta)
 = S(g_s, g, A, B) +   {\cal O}(\partial F).
\end{equation}

It was shown in \ct{yang} that the dual description through the exact SW map
is simply given by the identity \eq{equiv-dbi-gen} using
the change of variables between open and closed
string parameters, \eq{op-cl-gen} and \eq{Gs-gs-gen}.
More precisely, the dual description of the NC DBI action \eq{dbi-gen}
via the exact SW map is given by the ordinary one \eq{dbi-c}
expressed in terms of open string variables
\begin{eqnarray}\label{sw-equiv}
 &&  \int d^{p+1} x \sqrt{-\det(G + \kappa (\widehat{F}+ \Phi))} \nonumber \\
 && \hspace{3cm} = \int d^{p+1} x \sqrt{\det{(1+ F \theta})}
 \sqrt{-\det{(G + \kappa (\Phi + {\bf F}))}} +   {\cal O}(\partial F),
\end{eqnarray}
where
\begin{equation}\label{f-f}
    {\bf F} = \frac{1}{1 + F\theta} F.
\end{equation}
Note that the action \eq{sw-equiv} is exactly the same as
the DBI action obtained using the $\zeta$-function regularization
scheme by Andreev and Dorn \ct{andreev} (their Eq. (2.24)).

In the zero slope limit $\kappa \to 0$, Eq. (\ref{sw-equiv})
for $\Phi=0$ and $p=3$ defines an exact nonlinear action
of the SW deformed electrodynamics:
\begin{equation}\label{sw-flat}
- \frac{1}{4g_{YM}^2} \int d^4 x \widehat{F}_{\mu\nu}
\star   \widehat{F}^{\mu\nu} = - \frac{1}{4 g_{YM}^2}
\int d^4 x \sqrt{-\det{{\rm g}}} \;
{\rm g}^{\mu \alpha} {\rm g}^{\beta\nu} F_{\mu\nu}
F_{\alpha\beta},
\end{equation}
where we introduced an effective non-symmetric ``metric" induced
by the dynamical gauge fields such that
\begin{equation}\label{ind-metric-flat}
    {\rm g}_{\mu\nu} = \eta_{\mu\nu} + (F\theta)_{\mu\nu},
    \qquad  {\rm g}^{\mu\nu} =
    \Bigl(\frac{1}{\eta + F\theta}\Bigr)^{\mu\nu}.
\end{equation}
The NC Maxwell action after the SW map looks like the ordinary
Maxwell theory coupled to the ``induced metric" $\mathrm{g}_{\mu\nu}$.
It should be remarked that the gravitational field in the action \eq{sw-flat}
cannot be interpreted just as a fixed background since it depends
on the dynamical gauge field.
The identity \eq{sw-flat} is very remarkable in the sense
that the NC Maxwell action after the exact SW map can be regarded
as an ordinary field theory coupling to a field
dependent gravitational background \ct{rivelles}.

A simple yet nontrivial application of the mapping (\ref{sw-flat})
is in the context of conformal anomalies. The planar part of the conformal
anomaly in NC gauge theory is known \cite{nakajima} to be proportional
to the left hand side of Eq. (\ref{sw-flat}). Then using the map it is feasible to express
the result in terms of commutative variables. In this way the
conformal anomalies in the NC and commutative descriptions get
related. We might recall that current (divergence) anomalies in NC
and commutative theories are also related by appropriate SW maps \cite{banerjee}.

Another interesting case arises from the choice $\Phi_{\mu\nu} = - B_{\mu\nu}$,
which naturally appears in the matrix model \ct{sw,seiberg}. In this
case, with the Euclidean signature,
\begin{equation}\label{matrix}
    \theta = \frac{1}{B}, \qquad G= - \kappa^2 B \frac{1}{g} B,
    \qquad G_s = g_s \sqrt{\det(-\kappa B g^{-1})}
\end{equation}
and
\begin{equation}\label{matrix-f}
    (\widehat{F} + \Phi)_{\mu\nu} = i B_{\mu \lambda} [X^\lambda, X^\sigma]_\star B_{\sigma
    \nu},
\end{equation}
where
\begin{equation}\label{X}
    X^\mu = x^\mu + \theta^{\mu\nu} \widehat{A}_\nu.
\end{equation}
The DBI action related to the matrix model has more natural
description, so called, background independent formulation, in
terms of closed string variables \ct{seiberg}. The NC DBI action \eq{dbi-gen}
can be expressed instead in terms of closed string variables using
the relation \eq{matrix} and then the equivalence \eq{equiv-dbi-gen} defines the exact inverse
SW map
\bea \label{sw-matrix}
&& \frac{2\pi}{g_s (2\pi \kappa)^{\frac{p+1}{2}}}\int d^{p+1} x
\sqrt{\det(g + \kappa {\cal F})} \xx
&& \quad = \frac{2\pi}{g_s (2\pi \kappa)^{\frac{p+1}{2}}}\int d^{p+1} x
\sqrt{\det(1 - \theta \widehat{F})} \sqrt{\det(g + \kappa (B + \widehat{{\bf F}}))},
\eea
where
\begin{equation}\label{matrix-ncf}
    \widehat{{\bf F}} = \widehat{F}\frac{1}{1-\theta\widehat{F}}.
\end{equation}
Our result \eq{sw-matrix}
is consistent with the exact SW map obtained
by completely independent way in \ct{liu,okawa-ooguri,mukhi-sury,liu-mich}
as shown in next section.

Note that
\begin{equation}\label{matrix-bf}
 B  + \widehat{{\bf F}} = B \frac{1}{1 - \theta \widehat{F}}.
\end{equation}
In the zero slope limit, $\kappa \to 0$, now keeping fixed $g_{\mu\nu}$
and $g_{YM}^2$, we obtain an intriguing identity
\begin{equation}\label{sw-matrix-zero}
\frac{1}{4g_{YM}^2}\int d^{p+1} x {\cal F}_{\mu\nu} {\cal F}^{\mu\nu} =
\frac{1}{4g_{YM}^2}\int d^{p+1} x \sqrt{\det{\widehat{{\rm g}}}} \;
\widehat{{\rm g}}^{\mu \alpha} \widehat{{\rm g}}^{\beta\nu} B_{\mu\nu}
B_{\alpha\beta},
\end{equation}
where
\be \la{matrix-metric}
\widehat{{\rm g}}_{\mu \nu} = \delta_{\mu\nu} - (\theta
\widehat{F})_{\mu\nu}, \qquad \widehat{{\rm g}}^{\mu \nu} = \Bigl(\frac{1}{1 - \theta
\widehat{F}}\Bigr)^{\mu\nu}.
\ee
The identity \eq{sw-matrix-zero} definitely shows that
fluctuations $F$ with respect to the background $B$ induce
fluctuations of (NC) geometry from matrix model side.

\section{Geometric Construction of Exact Seiberg-Witten Map}

In this section we will demonstrate that the results in the
previous section can be reproduced in the context of deformation
quantization \ct{kon}. First let us start with a brief recapitulation
of the results in \ct{cornalba,jurco}.
Consider two symplectic forms
\begin{equation}\label{two-symp}
\omega_{\mu\nu}  = (\theta^{-1})_{\mu\nu} + F_{\mu\nu}, \qquad
B_{\mu\nu} = (\theta^{-1})_{\mu\nu}
\end{equation}
where $\theta^{\mu\nu}$ is a constant anti-symmetric tensor and
$F=dA$ is the field strength of Abelian gauge field $A_\mu$.
We shall assume that both $\omega$ and $B$ are non-degenerate.
We can associate star products $\star_\omega$ and $\star_B$ with
$\omega$ and $B$, respectively, in the context of the deformation quantization a
la Kontsevich \ct{kon}. The SW map is expressed in
terms of a transformation which relates the star product
associated with $B$ to the one associated with $\omega$.

Since $B$ and $\omega$ differ by an exact form, it is possible to
find a coordinate transformation $\rho$ which maps $\omega$ to
$B$, i.e., $\rho: x \to y= y(x)$ so that
\begin{equation}\label{darboux}
    \frac{\partial y^\alpha}{\partial x^\mu} \frac{\partial y^\beta}{\partial
    x^\nu} \omega_{\alpha\beta}(y) = B_{\mu\nu}.
\end{equation}
Thus the symplectic structures defined by $\omega$ and $B$ belong
to the same equivalence class and the two star products $\star_\omega$ and $\star_B$
must be equivalent. (Eq. \eq{darboux} is essentially the statement
of Darboux theorem on symplectic manifolds. We refer Sects. 3.2 and 3.3 in
\ct{mechanics} for a proof of this theorem and related symplectic
geometry.) In other words, we can eliminate any fluctuation of the
electromagnetic field strength by a simple coordinate
redefinition. More explicitly, there exists a map ${\cal D}$
acting on the space of functions which satisfies
\begin{equation}\label{k-map}
    {\cal D}(f \star_\omega g) = {\cal D}f \star_B {\cal D}g.
\end{equation}
In particular the NC gauge field is defined by
\begin{equation}\label{nc-a}
    X^\mu(x) = {\cal D}y^\mu \equiv x^\mu + \theta^{\mu\nu}
    \widehat{A}_\nu.
\end{equation}

This change of coordinates is not unique, but is defined up to
diffeomorphisms, the canonical transformations or
symplectomorphisms, which preserve $\theta$. The group of such diffeomorphisms is
non-Abelian and is generated by Hamiltonian vector fields of the form
$\delta x^\mu = \{x^\mu, S \}_\theta $ for some generating function $S$.
These diffeomorphisms replace the ordinary Abelian gauge
invariance of the original theory. That is, the NC gauge group is
the set of diffeomorphisms which leaves the two-form $B$
invariant. Within the framework of Kontsevich's deformation
quantization, the equivalence classes of Poisson manifolds can
thus be naturally identified with the sets of gauge
equivalence classes of star products on a Poisson manifold $M$ \ct{jurco},
which leads to the SW transformations. An important lesson from
the above arguments is that,
in the new coordinate system $y^\mu$, the dynamics is not
described by the $U(1)$ gauge potential $A_\mu (x)$, but is
described by the embedding functions $x^\mu(y)$ which are now the
dynamical fields \ct{cornalba}. In this way any fluctuation of the field
strength can be eliminated in favor of fluctuations of the induced
metric. This explains why the gauge fields play a role of gravity.

In Eq. \eq{k-map}, setting $f(y)= y^\mu$ and $g(y)= y^\nu$ and
supposing that the product between functions in the left and right
hand sides is defined by star products $\star_\omega$ and $\star_B$,
respectively, we get
\begin{equation}\label{def-sw}
    (\omega^{-1})^{\mu\nu}(X(x)) = (\theta - \theta \widehat{F}
    \theta)^{\mu\nu}(x)
\end{equation}
where the NC field strength $\widehat{F}_{\mu\nu}$ is given by Eq.
\eq{nc-f}. Eqs. \eq{two-symp} and \eq{def-sw} then lead to
\ct{liu}
\begin{equation}\label{iswmap}
    F_{\mu\nu}(X) = \Bigl(\widehat{F}\frac{1}{1-\theta
    \widehat{F}} \Bigr)_{\mu\nu}(x)
\end{equation}
or its inverse
\begin{equation}\label{swmap}
    \widehat{F}_{\mu\nu}(x) = \Bigl(\frac{1}{1 + F\theta} F
    \Bigr)_{\mu\nu}(X).
\end{equation}
Note that
\begin{equation}\label{anti-symm}
\widehat{F}\frac{1}{1-\theta \widehat{F}} = \frac{1}{1- \widehat{F}\theta}
\Bigl(\widehat{F}- \widehat{F}\theta \widehat{F} \Bigr) \frac{1}{1-\theta
    \widehat{F}} = \frac{1}{1- \widehat{F}\theta} \widehat{F}
\end{equation}
and similarly for Eq. \eq{swmap}. Antisymmetricity  of
$F_{\mu\nu}$ and $\widehat{F}_{\mu\nu}$ is guaranteed due to this
property.

Since
\begin{equation}\label{fourier1}
 F_{\mu\nu}(k) = \int d^{p+1} X F_{\mu\nu}(X) e^{i k \cdot X},
\end{equation}
we obtain
\begin{equation}\label{fourier2}
 F_{\mu\nu}(k) = \int d^{p+1} x \sqrt{\det(1-\theta \widehat{F})}
 \Bigl(\widehat{F}\frac{1}{1-\theta
    \widehat{F}} \Bigr)_{\mu\nu}(x) e^{i k \cdot X},
\end{equation}
where we used the formula
\be \la{volume}
d^{p+1} X = d^{p+1} x \sqrt{\det(1-\theta \widehat{F})}
\ee
which can be derived from Eqs. \eq{darboux} and \eq{def-sw}.
Note also that it follows from Eq. \eq{def-sw} that
\begin{equation}\label{determinant}
    \sqrt{\det(1-\theta \widehat{F}(x))}\sqrt{\det(1 + F(X)
    \theta)}= 1.
\end{equation}
Using the formula \ct{owl-identity}
\bea \la{wilson}
e^{i k \cdot X} &=& P_\star \exp\Bigl(i \int_0^1 d \tau
\p_\tau \xi^\mu(\tau) \widehat{A}_\mu(x + \xi(\tau)) \Bigr)
\star e^{i k \cdot x}, \xx
&=& W(x, C_k)\star e^{i k \cdot x},
\eea
where $P_\star$ denotes path ordering with respect to the $\star$-product
and $W(x, C_k)$ is a straight open Wilson line with path
$C_k$ parameterized by
\begin{equation}\label{path}
\xi^\mu(\tau) = \theta^{\mu\nu} k_\nu \tau,
\end{equation}
we can get the exact (inverse) SW map for the
field strength \ct{liu,okawa-ooguri,mukhi-sury,liu-mich}
\begin{equation}\label{esw-f}
 F_{\mu\nu}(k) = \int d^{p+1} x \sqrt{\det(1-\theta \widehat{F})}
 \Bigl(\widehat{F}\frac{1}{1-\theta
    \widehat{F}} \Bigr)_{\mu\nu}(x) W(x, C_k) \star e^{i k \cdot x}.
\end{equation}

It is easy to demonstrate the equivalence of the DBI actions, Eq.
\eq{equiv-dbi-gen}, using Eqs. \eq{op-cl-gen}, \eq{Gs-gs-gen}, \eq{iswmap}, \eq{swmap},
\eq{volume}, and \eq{determinant} \ct{liu,schupp}:
\bea \label{proof-sw}
&&   \frac{2\pi}{g_s (2\pi \kappa)^{\frac{p+1}{2}}}\int d^{p+1} X
\sqrt{-\det(g + \kappa (F(X) + B))} \xx
&& \hspace{2cm} = \frac{2\pi}{G_s
(2\pi \kappa)^{\frac{p+1}{2}}}\int d^{p+1} x
\sqrt{-\det(G + \kappa (\widehat{F}(x)+ \Phi))}.
\eea
Here we used different coordinates, $X^\mu$ and $x^\mu$, for
commutative and NC descriptions, respectively. There is a simple
reason for the use of these coordinates. A natural coordinate
system in the commutative description is $\theta$-independent, i.e. background independent,
one, which is $X$-coordinates \ct{seiberg}, while that in the NC one is
$x$-coordinates due to their simple commutation relation \eq{nc-space}.
It is also very easy to prove the exact SW maps, Eqs. \eq{sw-equiv}
and \eq{sw-matrix}, using Eqs. \eq{iswmap}, \eq{swmap},
\eq{volume}, and \eq{determinant}.

We here discuss the SW map for constant field strength. Note that
the arguments in this section should also hold
for this case. Thus we see that the exact SW maps, Eqs. \eq{sw-equiv}
and \eq{sw-matrix}, must be true even for the constant field
strength. If we thus take the solution \eq{swmap} (which is exact for
constant fields) plus the measure factor coming from Eq. \eq{volume},
then a simple correspondence immediately leads to Eq. \eq{sw-flat}, thereby
proving the identity for constant fields exactly. The measure
change thus correctly accounts for those terms which would
otherwise be dropped in the constant field approximation. This is
in conformity with similar observations \cite{liu}
needed to show the equivalence of DBI actions.

\section{Exact Seiberg-Witten Map for Scalar Fields}

In this section we will find an exact SW map for scalar fields in
the adjoint representation of gauge group. To get it, we will
apply the standard dimensional reduction scheme for the exact SW map,
Eq. \eq{iswmap} or \eq{swmap}. For this purpose, we compactify
one of the spatial directions and we denote its compact coordinate
as $z \in {\bf S}^1$ and the compact gauge fields along ${\bf S}^1$
as $\widehat{A}_z = \widehat{\varphi}$ and $A_z = \varphi$.
According to the usual dimensional reduction
scheme, we set $\theta^{\mu z} = - \theta^{z \mu} = 0$ where $\mu$
spans non-compact NC directions.

Adopting the standard rule, we identify
\bea \la{dim-red-nc}
\widehat{F}_{\mu z} (x) &=&  \partial_\mu \widehat{\varphi} -
i [\widehat{A}_\mu, \widehat{\varphi}]_\star \xx
&=& \widehat{D}_\mu \star\widehat{\varphi}(x), \\
\la{dim-red-c}
F_{\mu z} (X) &=&  \frac{\partial \varphi(X)}{\partial X^\mu}.
\eea
For the exact SW map \eq{swmap}, we take the ordering
\begin{eqnarray}\label{esw-scalar-cov}
\widehat{D}_\mu \star\widehat{\varphi}(x) &=& \widehat{F}_{\mu z}(x) \xx
&=& \Bigl(\frac{1}{1 + F\theta} \Bigr)_\mu^{\;\;\nu}
(X) \frac{\partial \varphi(X)}{\partial X^\nu}
\end{eqnarray}
since
\begin{equation}\label{esw-wrong-cov}
\widehat{F}_{z \mu}(x) = \Bigl(\frac{F \theta}{1 + F \theta} \Bigr)_\mu^{\;\;\nu}(X)
\frac{\partial \varphi(X)}{\partial X^\nu}
\end{equation}
does not produce the correct commutative limit when $\theta \to 0$.
We regard Eq. \eq{esw-scalar-cov} as the exact SW map for the
adjoint scalar field $\widehat{\varphi}$. It is straightforward to
generalize to the case being several adjoint scalar fields $\widehat{\varphi}_i, \;
i=1, \cdots,n$, by considering a similar dimensional reduction onto
${\bf T}^n$.

Now we will show that the SW map \eq{esw-scalar-cov} obtained by the dimensional
reduction scheme is consistent
with that obtained by studying RR couplings of unstable non-BPS
D-branes \ct{mukhi-sury}. Consider the coupling of a non-BPS
$Dp$-brane to the RR form $C^{(p)}$ in the commutative description
\bea \la{RR-coupling}
\int dT \wedge C^{(p)} &=& \int d^{p+1} X \varepsilon^{\mu_1 \mu_2
\cdots \mu_{p+1}} {\cal O}_{\mu_1}(X)
C^{(p)}_{\mu_2 \cdots \mu_{p+1}}(X) \xx
&=& \int d^{p+1} k \varepsilon^{\mu_1 \mu_2
\cdots \mu_{p+1}} \widetilde{{\cal O}}_{\mu_1}(k)
\widetilde{C}^{(p)}_{\mu_2 \cdots \mu_{p+1}}(-k),
\eea
where $T$ is the tachyon field and
\begin{equation}\label{tachyon-T}
{\cal O}_{\mu}(X) = \frac{\p T(X)}{\p X^\mu}.
\end{equation}
The same RR coupling $C^{(p)}$ of a NC non-BPS $Dp$-brane has the
following form for each momentum mode \ct{mukhi-sury}
\bea \la{nc-RR-coupling}
\varepsilon^{\mu_1 \mu_2
\cdots \mu_{p+1}} \widetilde{C}^{(p)}_{\mu_2 \cdots \mu_{p+1}}(-k)
\int \frac{d^{p+1} x}{(2\pi)^{p+1}} L_\star \Bigl[
\sqrt{\det(1-\theta\widehat{F})} \widehat{{\cal O}}_{\mu_1}(x)
W(x,C_k) \Bigr]\star e^{ik \cdot x}
\eea
where
\begin{equation}\label{nc-tachyon-T}
 \widehat{{\cal O}}_{\mu}(x)=
 \Bigl(\frac{1}{1 - \widehat{F}\theta} \Bigr)_\mu^{\;\;\nu}(x)
(\widehat{D}_\nu \star\widehat{T})(x).
\end{equation}

From the equivalence of commutative and NC couplings to the RR
form $C^{(p)}$ of a non-BPS $Dp$-brane, we find that
\begin{equation}\label{sw-tachyon}
    \widetilde{{\cal O}}_{\mu}(k)=\int \frac{d^{p+1} x}{(2\pi)^{p+1}} L_\star \Bigl[
\sqrt{\det(1-\theta\widehat{F})} \widehat{{\cal O}}_{\mu}(x)
W(x,C_k) \Bigr]\star e^{ik \cdot x}.
\end{equation}
In the DBI approximation, we get
\begin{equation}\label{isw-scalar}
    \frac{\p T(X)}{\p X^\mu} =
    \Bigl(\frac{1}{1 - \widehat{F}\theta} \Bigr)_\mu^{\;\;\nu}(x)
(\widehat{D}_\nu \star\widehat{T})(x)
\end{equation}
where we used Eqs. \eq{volume} and \eq{wilson}. This SW map can
simply be obtained from Eq. \eq{iswmap} by the dimensional reduction \eq{dim-red-nc}
and \eq{dim-red-c}, thus proving the consistency of our scheme.

For a real scalar field in the adjoint representation of $U(1)$,
the flat spacetime action for the NC scalar field is
\begin{equation}\label{action-scalar}
    \widehat{S}_{\widehat{\varphi}} = \frac{1}{2} \int d^4 x \widehat{D}^\mu
    \widehat{\varphi}\star \widehat{D}_\mu \widehat{\varphi}.
\end{equation}
The action is invariant under the gauge transformation
\begin{equation}\label{gauge-tr}
    \widehat{\delta}_{\widehat{\lambda}} \widehat{A}_\mu
    = \widehat{D}_\mu \star \widehat{\lambda},
    \qquad \widehat{\delta}_{\widehat{\lambda}}\widehat{\varphi}
    = -i [\widehat{\varphi}, \widehat{\lambda}]_\star.
\end{equation}
We apply the exact SW map \eq{esw-scalar-cov} to the action
(\ref{action-scalar}) and the result is
\begin{equation}\label{esw-scalar}
    \frac{1}{2} \int d^4 x \widehat{D}^\mu
    \widehat{\varphi}\star \widehat{D}_\mu
    \widehat{\varphi} = \frac{1}{2} \int d^4 x \sqrt{\det(1+ F\theta)}
\Bigl(\frac{1}{1+ F\theta}\frac{1}{1+ \theta F} \Bigr)^{\mu\nu}
    \partial_\mu \varphi \partial_\nu \varphi.
\end{equation}
Here we are using the same symbol $x$ to denote both the
commutative (the right hand side) and the NC (the left hand side)
coordinates. We will often use the symbol $x$ for both
descriptions when the distinction is not necessary. It can be
easily checked that the leading order in $\theta$ in the right hand side of
Eq. (\ref{esw-scalar}) exactly coincides with Eq. (7) in \cite{rivelles}.

The final form (\ref{esw-scalar}) after the exact
SW map can be recast to the form coupled to a gravitational
background with a specific dilaton coupling:
\be \la{ind-gravity-scalar}
S_{\varphi} = \frac{1}{2} \int d^4 x e^{-\phi} \sqrt{\det{\rm g}}
\;{\rm g}^{\mu\nu} \partial_\mu \varphi \partial_\nu \varphi,
\ee
where we introduced an induced symmetric metric
\be \la{ind-metric-scalar}
{\rm g}_{\mu\nu}=\Bigl((1+ \theta F)(1+ F\theta) \Bigr)_{\mu\nu},
\qquad
{\rm g}^{\mu\nu}=\Bigl(\frac{1}{1+ F\theta}\frac{1}{1+ \theta F} \Bigr)^{\mu\nu}
\ee
and a dilaton given by
\be \la{dilaton}
\phi = \frac{1}{4}\Tr \ln {\rm g}.
\ee
Our metric coupling \eq{ind-gravity-scalar} is different from
Eq. (8) in  \cite{rivelles}. (It is not possible to have the symmetric traceless metric
such as Eq. (9) in  \cite{rivelles} beyond the leading order in $\theta$.)
The metric in Einstein relativity is a property of spacetime
itself rather than a field over spacetime and thus all
non-gravitational fields should couple in the same manner to a
single gravitational field, sometimes called ``universal
coupling''. We here see that the metric coupling induced by
noncommutativity is not universal, i.e., species dependent.
However, this is somewhat expected since, in NC field theory
context, there does not exist a principle to guarantee the
universal coupling such as the Equivalence principle in Einstein
relativity. Although we could not yet find the exact SW map for a
scalar field in the fundamental representation of gauge group, we
think that the same thing also happens for that case.

\section{Topological Invariants and Exact Seiberg-Witten Map}

The coupling of D-branes to RR potentials \ct{liu,okawa-ooguri,mukhi-sury,liu-mich} is given by
\begin{equation}\label{wz-rr}
    S_{WZ} = \int d^{p+1} k Q(k) D(-k)
\end{equation}
where $D = C e^{\frac{\kappa}{2 \pi}B}$ and
\begin{equation}\label{qk}
    Q(k) = \int d^{p+1} x L_\star
    \Bigl[ \sqrt{\det(1-\theta \widehat{F})} e^{\frac{\kappa}{2 \pi}\widehat{F}
    \frac{1}{1-\theta \widehat{F}}} W(x, C_k) \Bigr]
    \star e^{i k\cdot x}.
\end{equation}
$Q(k=0)$ defines the charges of lower dimensional branes which can
be identified with a K-theory class $\mu(E)$ of a projective
module $E$ as an element of integral even cohomology \ct{liu-mich}. One can
immediately see that $Q(k=0)$ in terms of commutative coordinates
$X$
maps to the (integrated) Chern character ${\rm ch}(E)$
by Eqs. \eq{iswmap} and \eq{volume}, i.e.,
\bea \la{k-chern}
\mu(E) &=& \int d^{p+1} x \sqrt{\det(1-\theta \widehat{F})}
e^{\frac{\kappa}{2 \pi}\widehat{F}
    \frac{1}{1-\theta \widehat{F}}} \xx
&=& \int d^{p+1} X e^{\frac{\kappa}{2 \pi}F(X)} \xx
&=& \int d^{p+1} X {\rm ch}(E).
\eea
The identity \eq{k-chern} directly proves that the K-theory class $\mu(E)$ of a projective
module $E$ takes values in an integral even cohomology class. We
should emphasize that the whole argument in this section is
equally valid even for non-Abelian case although we present only
Abelian case for simplicity.

The identity \eq{k-chern} for four
dimensions is related to the topological charge of instantons
\be \la{map-instanton}
\frac{1}{64 \pi^2} \int d^4 x \sqrt{\det(1-\theta \widehat{F})}
\Bigl(\widehat{F}
    \frac{1}{1-\theta \widehat{F}} \Bigr) \wedge \Bigl(\widehat{F}
    \frac{1}{1-\theta \widehat{F}} \Bigr)
= \frac{1}{64 \pi^2} \int d^4 X F \wedge F
\ee
where we used the (star) wedge notation
\begin{equation}
    \widehat{F} \wedge \cdots \wedge \widehat{F} =
    \varepsilon^{\mu\nu \cdots \lambda\rho} \widehat{F}_{\mu\nu}
    \star \cdots \star \widehat{F}_{\lambda\rho}. \nonumber
\end{equation}
Related to the instanton number in NC gauge theory, the quantity
one usually calculates has the form instead
\be \la{instanton}
\frac{1}{64 \pi^2} \int d^4 x \widehat{F}  \wedge \widehat{F}
\ee
and it has been known \ct{instanton-number} that it is also integer valued. So it is
natural to expect that
\be \la{conjecture-instanton}
\int d^4 x \sqrt{\det\widehat{{\rm g}}}
\Bigl(\widehat{F}\widehat{{\rm g}}^{-1} \Bigr) \wedge
\Bigl(\widehat{F} \widehat{{\rm g}}^{-1}\Bigr)
=  \int d^4 x \widehat{F}  \wedge \widehat{F}
\ee
with the induced metric \eq{matrix-metric}.
We will prove the identity \eq{conjecture-instanton} on a more general ground.
First note that the $\theta$ dependence in $\mu(E)$ comes from the explicit
dependence on $\theta$ as well as the implicit one through the
definition of $\widehat{F}$ and the $\star$-product between them.
However, since we are working in the limit of slowly varying
approximation which is constantly assumed in the derivation of DBI
actions in string theory, we can ignore all derivatives of $\widehat{F}$
and regard the products in the expansion of $\mu(E)$ as ordinary
products, i.e. the implicit $\theta$ dependence is only in the definition
of $\widehat{F}$ \ct{sw}. To correctly incorporate the derivatives
of $\widehat{F}$, it is necessary to systematically include higher
order $\alpha^\prime$ corrections to both descriptions. Indeed it
was known \ct{wyllard} that there is such an $\alpha^\prime$ correction in
\eq{map-instanton}.

Now we will show that the explicit $\theta$ dependence in $\mu(E)$
actually vanishes in the approximation of neglecting derivatives
of $\widehat{F}$. Taking the derivatives with respect to the explicit
dependence, we get
\begin{equation}\label{theta-der}
    \delta_\theta \mu(E) = \int d^{p+1} x \sqrt{\det(1-\theta \widehat{F})}
\Bigl( - \Tr(\delta \theta \widehat{{\bf F}}) e^{\frac{\kappa}{2 \pi}\widehat{{\bf
F}}} + \frac{\kappa}{\pi}(\widehat{{\bf F}} \delta \theta \widehat{{\bf F}})
\wedge e^{\frac{\kappa}{2 \pi}\widehat{{\bf F}}}
\Bigr) = 0
\end{equation}
where we used the identity \ct{rabin2},
\begin{equation} \la{identity-f}
\theta^{\alpha\beta}F_{\alpha\beta}(F \wedge
\cdots \wedge F)_{{\rm n-fold}} = -2n (F\theta F)
\wedge (F \wedge \cdots \wedge F),
\end{equation}
which is valid for any antisymmetric tensors. This means that
there is no explicit dependence on $\theta$, so that $\mu(E)$ is
more simplified by setting $\theta =0$ whenever it occurs
explicitly. That is, as was first shown in \ct{rabin2},
\bea \la{elliott-nc}
\mu(E)&=&  \int d^{p+1} x \sqrt{\det(1-\theta \widehat{F})}
e^{\frac{\kappa}{2 \pi}\widehat{{\bf F}}} \xx
&=& \int d^{p+1} x e^{\frac{\kappa}{2 \pi}\widehat{F}(x)}.
\eea
This proves the identity \eq{conjecture-instanton} for $p=3$.
As a simple corollary, we also get
\bea \la{elliott-c}
&&  \int d^{p+1} X \sqrt{\det(1+ F(X)\theta)}
e^{\frac{\kappa}{2 \pi}{\bf F}(X)} \xx
&=& \int d^{p+1} X e^{\frac{\kappa}{2 \pi}F(X)} = \int d^{p+1} X  {\rm ch}(E)
\eea
In particular,
\be \la{conjecture-c}
\int d^4 x \sqrt{\det{\rm g}}
\Bigl({\rm g}^{-1} F \Bigr) \wedge \Bigl({\rm g}^{-1} F\Bigr)
=  \int d^4 x F  \wedge F.
\ee
Noting that the instanton action is topological, i.e. independent
of the background metric,
the above identity together with Eq. \eq{conjecture-instanton}
seems to perfectly agree with our interpretation on the induced
metric.

The integrated Chern character can be expressed in terms of
Chern-Simons form $\widehat{\Omega}$ such that
\begin{equation}\label{chern}
    \int d^{p+1} x (\widehat{F} \wedge \cdots \wedge
    \widehat{F})_{{\rm (p+1)-form}} = \int d^{p+1} x \; d \widehat{\Omega}_{p}
\end{equation}
with
\begin{equation}\label{chern-simons}
\widehat{\Omega}_{p} = \frac{p-1}{2} \int^1_0 dt (\widehat{A} \wedge
\widehat{F}_t \wedge \cdots \wedge \widehat{F}_t)
\end{equation}
where $\widehat{F}_t = t d\widehat{A} -i t^2 \widehat{A} \star \widehat{A}$.
The proof of Eq. \eq{chern} is essentially the same as the
ordinary non-Abelian case since the cyclic property is available
under the integral. The exact SW map \eq{k-chern} together with the identity
\eq{elliott-nc} implies that
\be \la{esw-cs}
\int d^{p} x \; \widehat{\Omega}_{p}(\widehat{A},
\widehat{F}_t)(x)
=\int d^{p} X \; \Omega_{p}(A,F_t)(X).
\ee

If we definitely take care of the star products between
$\widehat{F}$'s, the SW map \eq{esw-cs} may not be quite true.
For $p=3$, however, we have the following property
\begin{equation}\label{star-2}
    \int d^4 x (\widehat{f} \star \widehat{g})(x) =
    \int d^4 x (\widehat{f} \widehat{g})(x)
\end{equation}
and, using the SW map \eq{swmap} and the identity \eq{conjecture-c}, we get
\bea \la{esw-3cs}
\int d^4 x (\widehat{F} \wedge \widehat{F})(x) &=& \int d^4 X \sqrt{\det(1+ F(X)\theta)}
({\bf F} \wedge {\bf F})(X) \xx
&=& \int d^4 X (F \wedge F)(X).
\eea
The same result also directly follows by taking the variation
of the left side of Eq. \eq{esw-3cs} with respect to the NC parameter $\theta$
and showing that it vanishes \cite{rabin2}. Moreover the same map
can be obtained by exploiting the SW map for anomalous axial
currents \ct{banerjee}.
Therefore the identity \eq{esw-cs} is still true for $p=3$ as was proved
earlier in \ct{grandi}. It was also shown \ct{kaminsky} that the equivalence
persists at the quantum level in perturbation theory. For higher
dimensions, e.g. $p=5$, NC ordering effect comes in, i.e.,
\begin{equation}\label{star-3}
    \int d^4 x (\widehat{f} \star \widehat{g} \star \widehat{h})(x)
    \neq
    \int d^4 x (\widehat{f} \widehat{g} \widehat{h})(x).
\end{equation}
Thus we cannot simply replace $\widehat{F}(x)$ by ${\bf F}(X)$
due to the explicit derivatives of $\widehat{F}(x)$ in the
star products. The NC ordering effect essentially
spoils the property \eq{esw-cs}, the form invariance of the Chern-Simons action under
the SW map, as was shown by Polychronakos \ct{poly-cs}.

\section{Discussion}

In this paper we revisited the dual description on DBI actions
via the exact SW map recently obtained by one of us in \ct{yang}.
We showed that the deformation quantization scheme clearly
explains why the dual description via SW map includes a
fluctuating geometry induced by gauge fields and noncommutativity,
in a sense, reflects the presence of a fluctuating ``medium".
Furthermore the picture on the induced gravity has been
generalized to an adjoint scalar field and has been particularly
useful to understand topological invariants in NC field theories.
This picture may have many interesting implications in both string
theory and field theory.

Our discussions so far have been confined only to Abelian gauge
group and to the DBI limit. Thus several interesting open issues
remain for the future. First of all, it will be interesting to
find non-Abelian generalization \ct{na-dbi} and an exact SW map for DBI actions
with derivative corrections \ct{wyllard,der-correction}.
Let us discuss some issues briefly.

When one has $N$ coincident type II $Dp$-branes, the worldvolume
theory is a $U(N)$ gauge theory. The explicit construction of such
an action is a difficult problem that is not yet completely
settled. But one may simply adopt the non-Abelian DBI action
proposed by Tseytlin \ct{na-dbi} in terms of a symmetrized trace prescription over the
Chan-Paton indices. Fortunately Terashima already showed \ct{sw-nadbi}
(see also \ct{nadbi-cornalba}) the equivalence between the
non-Abelian DBI action and its NC counterpart in the approximation
of neglecting derivative terms, using the differential equation
defining the SW map and the change of variables \eq{op-cl-gen} and
\eq{Gs-gs-gen}, following a similar recipe to \ct{sw}.
In the rank one case, the exact SW map \eq{sw-equiv}
in the DBI approximation has been obtained
by simply applying the change of variables \eq{op-cl-gen} and
\eq{Gs-gs-gen} to the commutative DBI action \eq{dbi-c}.
Thus the exact SW map for the non-Abelian DBI action may also be
obtained similarly. It should be straightforward to check this
claim \ct{next}.

The exact SW map \eq{sw-flat} raises several interesting issues.
Apart from its implications for conformal anomalies already
stated, it should be useful in analyzing renormalizability, UV/IR
mixing, and unitarity of SW deformed electrodynamics. Also it
would be an interesting question as to whether or not the
equivalence implied by the SW map persists even in the
nonperturbative level. This question is rather subtle.
It was pointed out \ct{sw,seiberg} that
the change of variables from $\widehat{A}_\mu$ to $A_\mu$ or vice
versa has only a finite radius of convergence. This can be seen
from Eq. \eq{darboux} that when $\p y^\alpha/\p x^\mu$ has
a zero eigenvalue, $F$ must diverge. Similarly, when $\omega$ has
a zero eigenvalue, $\p y^\alpha / \p x^\mu$ must diverge. Thus
the SW map may not completely encode the topology of gauge fields.
One example may be the level quantization of NC Chern-Simons
theory for $U(1)$ gauge group \ct{nc-cs,poly-cs}. In Eq.
\eq{esw-cs}, we showed the equivalence between the commutative and
NC Chern-Simons theories. However, in the course of the
derivation, it was implicitly assumed that both $\p y^\alpha/\p x^\mu$ and
$\omega_{\mu\nu}$ are nonsingular, which is not completely true.
Thus the nonperturbative aspects of the SW map remain as an
interesting future problem.

NC soliton solutions have been studied via SW map in several
contexts \ct{soliton}. Along this direction, it will be
interesting to reexamine NC $U(1)$ instantons \ct{instanton}
in view of the SW deformed electrodynamics \eq{sw-flat}.
As Eq. \eq{esw-3cs} implies the rigidity of the topological charge
of instantons under the SW map, an instanton solution on NC space
must ensure the existence of the corresponding commutative
instanton. It may be explicitly checked by studying the
self-duality equations for both sides of Eq. \eq{sw-flat} which are related
to each other by the SW map \eq{swmap}. The correspondence between
the commutative and NC instantons seems to lead to an intriguing
picture that NC $U(1)$ instantons may be identified
with Abelian instantons on ALE space \ct{ale}. \\

{\bf Acknowledgment} We thank Nicolas Grandi for helpful comments.
HSY was supported by the Brain Korea 21 Project in 2004.

\newpage


\nc{\npb}[3]{Nucl. Phys. {\bf B#1} (#2) #3}

\nc{\plb}[3]{Phys. Lett. {\bf B#1} (#2) #3}

\nc{\prl}[3]{Phys. Rev. Lett. {\bf #1} (#2) #3}

\nc{\prd}[3]{Phys. Rev. {\bf D#1} (#2) #3}

\nc{\ap}[3]{Ann. Phys. {\bf #1} (#2) #3}

\nc{\prep}[3]{Phys. Rep. {\bf #1} (#2) #3}

\nc{\epj}[3]{Eur. Phys. J. {\bf #1} (#2) #3}

\nc{\ptp}[3]{Prog. Theor. Phys. {\bf #1} (#2) #3}

\nc{\rmp}[3]{Rev. Mod. Phys. {\bf #1} (#2) #3}

\nc{\cmp}[3]{Comm. Math. Phys. {\bf #1} (#2) #3}

\nc{\mpl}[3]{Mod. Phys. Lett. {\bf #1} (#2) #3}

\nc{\cqg}[3]{Class. Quant. Grav. {\bf #1} (#2) #3}

\nc{\jhep}[3]{J. High Energy Phys. {\bf #1} (#2) #3}

\nc{\atmp}[3]{Adv. Theor. Math. Phys. {\bf #1} (#2) #3}

\nc{\hepth}[1]{{\tt hep-th/{#1}}}


\end{document}